\documentclass[prl,amsmath,aps,showpacs,twocolumn,superscriptaddress]{revtex4}
\usepackage{graphicx,xspace}
\usepackage{float}
\usepackage{amsmath}
\usepackage{amssymb}
\usepackage[normalem]{ulem}

\input{epsf}

\usepackage{dcolumn}
\usepackage{bm}

\begin{document}

\title{Crossed ratchet effects for magnetic domain wall motion}

\author{A. P\'erez-Junquera}
\affiliation{Depto. F{\'i}sica, Universidad de Oviedo, 33007 Oviedo,
Spain}

\author{V. I. Marconi}
\affiliation{Depto. F{\'i}sica At\'omica, Molecular  y Nuclear, and  {\em GISC}, Universidad Complutense de Madrid, 28040 Madrid, Spain}

\author{A. B. Kolton}
\affiliation{Depto. F{\'i}sica At\'omica, Molecular  y Nuclear, and  {\em GISC}, Universidad Complutense de Madrid, 28040 Madrid, Spain}

\author{L. M. \'Alvarez-Prado}
\affiliation{Depto. F{\'i}sica,
Universidad de Oviedo, 33007 Oviedo, Spain}

\author{Y. Souche}
\affiliation{Institut N\'eel, CNRS/UJF, BP 166, 38042 Grenoble,
France}

\author{A. Alija}
\affiliation{Depto. F{\'i}sica, Universidad de Oviedo, 33007 Oviedo,
Spain}

\author{M. V\'elez}
\email{mvelez@uniovi.es}
\affiliation{Depto. F{\'i}sica, Universidad de
Oviedo, 33007 Oviedo, Spain}

\author{J. V. Anguita}
\affiliation{Instituto de Microlectr\'onica de Madrid (CNM-CSIC), Tres Cantos, 28760 Madrid, Spain}

\author{J. M. Alameda}
\affiliation{Depto. F{\'i}sica, Universidad de Oviedo, 33007 Oviedo,
Spain}

\author{J. I. Mart{\'i}n}
\affiliation{Depto. F{\'i}sica, Universidad de Oviedo, 33007 Oviedo,
Spain}

\author{J. M. R. Parrondo}
\affiliation{Depto. F{\'i}sica At\'omica, Molecular  y Nuclear, and  {\em GISC}, Universidad Complutense de Madrid, 28040 Madrid, Spain}

\begin{abstract}
We study both experimentally and theoretically the driven motion of
domain walls in extended amorphous magnetic films patterned with a
periodic array of asymmetric holes. We find two crossed ratchet
effects of opposite sign that change the preferred sense for domain
wall propagation, depending on whether a flat or a kinked wall is
moving. By solving numerically a simple $\phi^4$-model we show
that the essential physical ingredients for this effect are quite
generic and could be realized in other experimental systems
involving elastic interfaces moving in multidimensional
ratchet potentials.
\end{abstract}


\maketitle The propagation of domain walls in thin ferromagnetic
films is a problem of great current interest. It provides both the
basis for a wide number of modern magnetic devices~\cite{review}
and an excellent experimental system to study the basic physics of
an elastic interface in the presence of either ordered or random
pinning defects~
\cite{paredes1,paredes1c,paredes2,paredes3,paredes4}. Such physics
has been indeed recently considered in many other experimental
systems involving interfaces, such as ferroelectric domain
walls~\cite{ferroelectrico}, contact lines of liquids
menisci~\cite{contact_line} or fractures~\cite{cracks}.
Furthermore,  it is relevant for systems involving periodic
elastic manifolds, such as vortex lattices in
superconductors~\cite{vinokur}, charge density
waves~\cite{nattermann_cdw_review} or Wigner
crystals~\cite{giamarchi_electronic_crystals_review}.

A case of particular interest  appears when the pinning potential
is asymmetric, favoring the propagation of the elastic interface
in one direction. This gives rise to several ratchet
effects~\cite{ratchets_reviews}, which are a potential tool to
control motion at micro and nanoscales in a variety of
systems~\cite{vicent}. One of the first known examples of ratchet
potentials in the field of magnetism is the use of ``angelfish''
patterns for controlling the sense of propagation of bubble
domains in domain shift registers \cite{angelfish}. Much more
recently, the asymmetric motion of domain walls (DWs) in nanowires
with a triangular structure \cite{allwood} or with a set of
asymmetric notches \cite{Himeno,Nori} has also been reported. In
all previous cases, domain wall propagation is restricted to a
narrow 1D path (either by narrow guide rails or by the nanowire
geometry) and its transverse wandering can be neglected. Then the
wall behaves essentially as a point particle in a 1D asymmetric
potential, provided effectively by the 2D geometry of the
patterned film. However, in a thin extended film, a DW is an
elastic line that can distort all along its length in response to
the 2D asymmetric pinning potential. The competition between
elasticity and pinning is a purely collective behavior and can
thus yield novel ratchet phenomena.

In this work, we study the propagation of DWs in {\it extended}
amorphous magnetic films patterned with a periodic array of
asymmetric holes. We observe experimentally, for the first time, two
crossed ratchet effects of opposite sign that change the preferred
sense for DW motion depending on whether a flat or a kinked wall is
moving. By identifying the essential physics we show
that this effect could be realized in other {\it multidimensional}
ratchet systems involving the motion of elastic interfaces.

\begin{figure}[ht]
 \begin{center}
 \includegraphics[angle=0,width=0.9\linewidth]{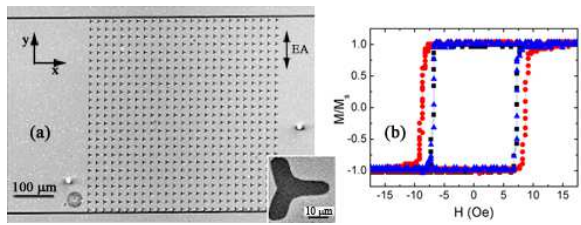}
 \end{center}
 \caption{(color online) (a) Scanning electron microscopy image of
 an array of asymmetric holes. Inset shows a detail of a single
 arrow hole. The easy axis direction is also indicated. (b) MOKE
 hysteresis loop measured at the array (circles) and at the
 unpatterned regions at the right (squares) and at the left
 (triangles) of the array.} \label{Fabricacion}
 \end{figure}

Amorphous 40 nm thick  magnetic Co-Si films have been fabricated by
sputtering with a well defined uniaxial anisotropy and a low
coercivity \cite{EPJB}. In these films, easy axis (EA) magnetization
reversal takes place by propagation of $180^\circ$ N\'eel walls that
tend to lie parallel to the EA \cite{JAP}. A $500 \times 500 \,\mu
{\rm m}^{2}$ ordered array of asymmetric antidots has been patterned
by a combination of e-beam lithography and an Ar$^{+}$ etching
process \cite{Nanotechnology}. Each hole is shaped as a small arrow
pointing perpendicular to the uniaxial EA (see
Fig.~\ref{Fabricacion}(a)). This allows us to define two different
senses of propagation for a DW lying along the EA (\emph{Y} axis):
``forward'' ($\mathcal{F}$), from left to right, i.e towards the
direction pointed by the arrows, and ``backward'' ($\mathcal{B}$),
from right to left. The asymmetric antidots are arranged in a square
array parallel to the EA, with a $20 \times 20 \,\mu {\rm m}^{2}$
unit cell, centered in  a $500 \,\mu {\rm m}$ wide path and
separated from the rest of the film by a $5\,\mu {\rm m}$ wide
trench.
Magnetic properties have been characterized both by Transverse
Magnetooptical Kerr effect (MOKE), using a setup with a laser
focused in a $300\,\mu {\rm m}$ spot in the desired sample
area~\cite{JAP}, and by MOKE microscopy~\cite{Souche}. The magnetic
field, $H$, is applied parallel to the film plane and along the EA.

\begin{figure}[ht]
\begin{center}
\includegraphics[angle=0,width=0.8\linewidth]{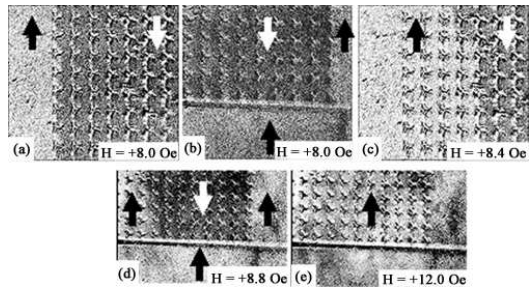}
\end{center}
\caption{(a)-(e) Sequence of MOKE microscopy images taken in the
ascending field branch of the hysteresis loop after saturation in
a negative field.} \label{microscopia}
\end{figure}

In Fig.~\ref{Fabricacion}(b) we show hysteresis loops measured both
at the array area (circles) and at the unpatterned regions in the
left (triangles) and in the right (squares) sides of the array. The
observed increment in coercive field ($H_{C}$) from $6.5\,$Oe in the
continuous film to $8.6\,$Oe
 in the array
is a  clear evidence that the fabricated holes act as effective
pinning centers, useful to control the DW motion. This difference in
coercivity implies that there is a field range, approximately
between $6.5\,$Oe and $8\,$Oe, where the continuous regions have
been reversed but not the patterned area, which will be bounded by
two DWs at its left and right sides. This is indeed observed in the
Kerr microscopy images shown in Figs.\ref{microscopia}(a-b) taken at
$H = 8\,$Oe after saturating the sample with a large negative field.
A DW can be identified in each image as the line separating the
dark-clear contrast regions (i.e. negative and positive
magnetization). The walls are located either at the first
(Fig.~\ref{microscopia}(a)) or at the last
(Fig.~\ref{microscopia}(b)) column of defects, indicating that they
cannot move further inside the
array area due to the antidot pinning.
Upon further increasing the applied field to $H=8.4\,$Oe, the left
wall penetrates the array, (see Fig.~\ref{microscopia}(c), in which
the left wall is pinned between the $4^{\rm th}$ and $5^{\rm th}$
antidot columns). Then, at $H=8.8\,$Oe, it can be seen in
Fig.~\ref{microscopia}(d) that the left wall has propagated in the
$\mathcal{F}$ direction, now up to the $17^{\rm th}$ defect column,
whereas the right wall has not been yet able to move. Finally, for
larger fields, both walls coalesce completing the magnetization
reversal at the array area (Fig.~\ref{microscopia}(e)).
This reversal sequence clearly shows that the depinning field for
$\mathcal{F}$ wall propagation $H_{\mathcal{F}} \approx 8.4\,$Oe is lower than
the field for $\mathcal{B}$ wall depinning $H_{\mathcal{B}} \geq 8.8\,$Oe (only
a lower bound can be obtained in this case) indicating that the
arrow-shaped holes act as asymmetric pinning centers for the DWs. A
similar image sequence for the descending field branch in the
hysteresis loops shows again a wall entering from the left and
moving in the $\mathcal{F}$ direction, only with an overall exchange
between the black/white contrast regions (i.e an overall
magnetization sign change). It is important to note that the easy
direction of motion in the patterned area is that in which the
length of the pinned wall between two antidots increases smoothly,
in agreement with the reported behavior in nanowires of triangular
cross section~\cite{allwood,Himeno}.

\begin{figure}[ht]
\begin{center}
\includegraphics[angle=0,width=0.9\linewidth]{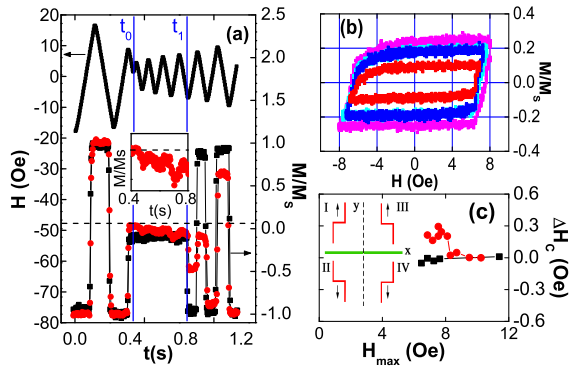}
\end{center}
\caption{(color online) (a) $H(t)$ sequence used to introduce a wall
inside the array and measure its propagation within it (top);
corresponding $M(t)$ response in the array (red circles) and in the
continuous film (black squares). Inset shows a zoom of the decrease in magnetization
from $t_0$ to $t_1$ in the array. (b) Minor hysteresis loops measured after
introducing a wall inside the array at the positive coercivity. (c)
Coercive field asymmetry as a function of minor loop amplitude in
the array (red symbols) and in the unpatterned area (black symbols).
Inset shows the results of X axis reflection and Y axis reflection
(broken symmetry in the array) on a kink moving upward.}
\label{ciclos}
\end{figure}

To study the motion of a single DW inside the array, the following
experiment has been performed (see the field \emph{vs.} time
sequence $H(t)$ in Fig.~\ref{ciclos}(a)): after a first complete
major loop between $t=0$ and $t=0.25$ s, the sample is saturated
in a large negative field. Next, the field is increased up to the
positive coercivity, where a single DW enters in the middle of the
array and, then, $H$ is decreased. Finally,
at $t=t_{0}$, a triangular field ramp of increasing amplitude
$H_{max}$ is applied to the sample. The corresponding time
evolution of the magnetization is shown at the bottom of
Fig.~\ref{ciclos}(a), both in the patterned and unpatterned
regions. Surprisingly, during several field cycles (between
$t_{0}$ and $t_{1}$), there is a net {\it decrease} in the
magnetization of the array (see inset of Fig.~\ref{ciclos}(a)),
indicating $\mathcal{B}$ DW motion. The sign of the $\partial
M/\partial t$ slope at $t=t_{0}$ is found to depend only on the
sign of the saturation magnetization $M_{S}$ before introducing
the wall inside the array, and not on the sign of $\partial
H/\partial t$, as would be in a standard accommodation
effect~\cite{accommodation}. Fig.~\ref{ciclos}(b) shows several
stable minor loops, measured with a similar $H(t)$ sequence as in
Fig.~\ref{ciclos}(a), but with a constant amplitude $H_{max} <
8\,$Oe in the triangular ramp after $t_{0}$ and centered along the
magnetization axis. These minor loops exhibit a clear asymmetry,
quantified by the difference between the positive and negative
coercivities $\Delta H_{C}= H_{C}^{asc}-H_{C}^{desc}$ of about
$0.2\,$Oe (Fig.~\ref{ciclos}(c)). Different from exchange bias,
the sign of $\Delta H_{C}$ is found to depend on the sign of
$M_{S}$ before introducing the wall, so that coercivity is always
lower when the wall is pushed in the $\mathcal{B}$ direction.

From these data, two main results are worth remarking:
first, the system ``keeps memory'' of the last saturating state that can be read
in the sign of $\Delta H_{C}$ or of $\partial M/\partial t$ at
$t=t_{0}$. Second, there is a clear change
between the behavior observed in Fig.~\ref{microscopia}, in which
the DW penetrates into the array more easily in the $\mathcal{F}$
direction, and the minor loop experiment of Fig.~\ref{ciclos},
in which DW motion within the array is
easier in the $\mathcal{B}$ direction.

\begin{figure}[ht]
\begin{center}
\includegraphics[angle=90,width=0.95\linewidth]{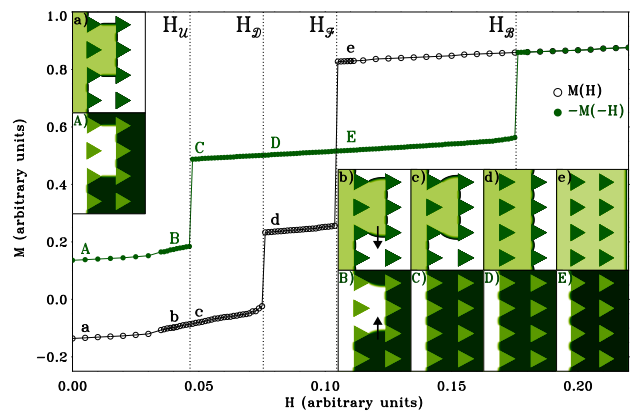}
\end{center}
\caption{(color online) Numerical results for the magnetic response of a kinked
domain wall vs  applied magnetic field. The initial state at $H=0$
is a wall with a kink-antikink pair (insets a) and A)), which
evolves asymmetrically with respect to $H$ and $-H$. Insets are
snapshots of the local magnetization $\phi$, for different pairs of
fields of equal magnitude, marked in the magnetization curves as
a,...e, in $M(H)$, and as A,...E in $-M(-H)$.
 Critical fields show clearly that it is easier
 to move a flat wall to the right
 than to the left but on the contrary the kinked
 wall is harder to move to the
 right than to the left.} \label{figura_phi4} \end{figure}

To understand these opposite effects, a crucial observation is
the change in the wall configuration as it enters the array: in the
continuous area, the wall is essentially flat (Fig.~\ref{microscopia}(a-b)) but
it develops kinks when it is pinned into the array
 (Fig.~\ref{microscopia}(c)). This suggests
an extra mechanism for DW motion in the minor loop experiment,
through upward/downward ($\mathcal{U}$/$\mathcal{D}$) kink
propagation, that is possible in our
geometry but not in the more restricted nanowire case where the DW
can not develop kinks. 
Indeed, by taking into account the reflection symmetries of the
array as depicted in the inset of Fig.~\ref{ciclos}(c), it is easy
to see that a kink propagating $\mathcal{U}$ (I) would be
equivalent to an antikink propagating $\mathcal{D}$ (II) but not
necessarily equivalent to a kink propagating $\mathcal{D}$ (IV) or
to an antikink propagating $\mathcal{U}$ (III) (see also the
insets in Fig.~\ref{figura_phi4}). Therefore, the array could in
principle induce an {\it asymmetric pinning for kink motion}, i.e.
perpendicular to the previously described ratchet affecting the
$\mathcal{F}$/$\mathcal{B}$ propagation of the flat wall. Most
interestingly, the experimental results of Figs.~\ref{microscopia}
and~\ref{ciclos} suggest that these {\it crossed ratchet effects}
must be of opposite sign.

To check the above scenario we consider the competition between
drive, elasticity and the asymmetric pinning on a single driven DW.
For this purpose we  simulated the paradigmatic $\phi^4$-model for a scalar order
parameter $\phi(x,y;t)$, in which a DW provides a smooth transition
between energetically equivalent minima of a simple free
energy~\cite{Chaikin}. We will show that this approach, although
simplistic as it avoids many of the complications of the full
micromagnetic model, has the main physical ingredients for the
effect, and most importantly, allows us to demonstrate the general
nature of the observed ratchet phenomena.
In our model $\phi(x,y;t)$ can be thought as a projection of the
coarse-grained magnetization vector along the easy direction. We
consider the evolution of $\phi$ in the domain
$\Omega-\bigtriangleup$, which includes all the space $\Omega$,
except the region $\bigtriangleup$ occupied by antidots. In order to
model the absence of magnetic material in $\bigtriangleup$, we set
Neumann boundary conditions $\partial_{\bf n} \phi|_{\partial
\bigtriangleup} = 0$ at the antidot borders ${\partial
\bigtriangleup}$.
Finally, considering a purely dissipative
dynamics, the equation of motion for $\phi$ can be written
as~\cite{Chaikin}
\begin{equation}
\eta \partial_t \phi = c \nabla^2 \phi + \epsilon_0 \biggl(\phi -
\phi^3\biggr) + H\label{eq_of_motion}
\end{equation}
where $c$ is the elastic stiffness of the order parameter,
$\epsilon_0$ is proportional to the local barrier separating the two equivalent minima of
the free energy density,
$H$ represents the magnetic field, and the friction coefficient
$\eta$ sets the microscopic time-scale. The relevant parameters $c$
and $\epsilon_0$ will determine both the width $\xi \propto
\sqrt{c/\epsilon_0}$ and the line tension $\sigma \propto \sqrt{c
\epsilon_0}$ of the DW~\cite{Chaikin}.

For the simulation we chose $\epsilon_0=\eta=1$ and $c$ such that
$\xi$ is $10\%$ of the characteristic size of the antidots, which
approximately corresponds to the realistic situation~\cite{JAP},
although we obtain qualitatively the same behavior for a finite
range of parameters. We solve numerically the
Eq.~(\ref{eq_of_motion}) in a $L_x \times L_y$
box with periodic boundary conditions in the $Y$ direction, and
model the asymmetrical antidots as a rectangular array of triangular
holes pointing to the positive $X$ direction (see insets in
Fig.~\ref{figura_phi4}). To ensure the presence of a DW along the
sample, we set $\phi(x=0,y;t)=1$ and $\phi(x=L_x,y;t)=-1$ as boundary
conditions in the X direction, and then probe the response of the DW
to different, positive and negative magnetic fields $H$. Since we
are interested in the response to constant or low-frequency fields,
we will only analyze the stationary magnetization $M(H)$,
starting with the particular initial condition of a single flat DW
with a kink-antikink pair (see insets A and a in
Fig.~\ref{figura_phi4}).

In Fig.~\ref{figura_phi4} we show the magnetization $M$ vs the
applied magnetic field $H$ starting at $H=0$ with the kink-antikink
pair \cite{epaps}. As indicated by the vertical dotted-lines we can clearly
distinguish four critical fields: $H_{\mathcal{U}}$ corresponds to the $\mathcal{U}$
($\mathcal{D}$) depinning of a kink (antikink), cases I and II in
Fig.~\ref{ciclos}(c), which amounts to a net motion of our initial
DW to the left ($\mathcal{B}$); $H_{\mathcal{D}}$ is the depinning field for
the $\mathcal{D}$
($\mathcal{U}$) motion of a kink (antikink), cases IV and III in
Fig.~\ref{ciclos}(c); $H_{\mathcal{F}}$ and $H_{\mathcal{B}}$
corresponding to the $\mathcal{F}$ and $\mathcal{B}$ depinning
fields of flat walls, respectively. We find that $H_{\mathcal{U}}
< H_{\mathcal{D}} < H_{\mathcal{F}} < H_{\mathcal{B}}$. As
expected, transport at low fields $|H|<H_{\mathcal{F}}$ is
dominated by the presence of mobile kinks. More interestingly, we
have $H_{\mathcal{U}} < H_{\mathcal{D}}$. This implies that a net
directed transport of the wall in the $\mathcal{B}$ direction can
be obtained under an low-frequency ac-field of amplitude
$H_{\mathcal{U}} < |H| < H_{\mathcal{D}}$, in qualitative
agreement with the behavior of the magnetization in the minor loop
experiment, between $t_0$ and $t_1$ shown in Fig.~\ref{ciclos}(a).
Finally, by increasing the magnetic field amplitude $|H|$ we have
an inversion of the rectification for flat walls, since
$H_{\mathcal{B}}> H_{\mathcal{F}}$, i.e., the wall as a whole
moves more easily in the $\mathcal{F}$ direction, as in the
experiments (cf. Fig.~\ref{microscopia}). The simulations also
show which are the key ingredients for the inversion in the
rectification between these two crossed ratchet effects: whereas
the $\mathcal{B}$ motion of a flat wall (Fig.~\ref{figura_phi4}
inset E) involves a sudden (i.e. long-range correlated) depinning
from the stable position at the base of the triangles, making it
the {\it hard} direction of motion, the $\mathcal{B}$ motion of a
kinked wall involves the $\mathcal{U}$ motion of a kink
(Fig.~\ref{figura_phi4} inset B), which gradually peels off the
wall from the triangle bases thus making this, on the contrary,
the {\it easy} direction of motion for a kinked wall at low
fields. It is worth noting that this behaviour is due to the
generic interplay between elasticity, pinning, and drive: while
the first two tend to minimize the line energy of the DW by
respectively straightening all segments and by optimally using the
holes bridging them, the applied field tends to increase the area
behind the DW with $\phi H > 0$. This leads to the characteristic
catenary shape of all pinned segments and finally to the
asymmetric depinning configurations and forces, responsible for
the observed ratchet effects.

In summary, our experimental and theoretical study of
the DW propagation across an array of asymmetric holes has revealed
the existence of two crossed ratchet effects: the first one favors
$\mathcal{F}$ motion of a flat wall while the second acts on the
$\mathcal{U}$/$\mathcal{D}$ kink propagation favoring net
$\mathcal{B}$ wall motion at low fields. As a result of the
interplay between both ratchets, the system keeps memory of the
sign of the last saturating state even in a zero magnetization
configuration, thus opening an interesting possibility for future
applications in memory devices. It is worth noting that this effect
relies completely in the extended nature of the DW, which allows
excitations transverse to the direction of propagation. Moreover,
the identification of the main physical ingredients for this novel
effect shows that it could be realized in other experimental systems
involving the motion of elastic interfaces or domain walls in
multidimensional ratchet potentials.

Work supported by Spanish CICYT under grants NAN2004-09087, MOSAICO,
and FIS2005-07392. We acknowledge helpful discussions with Prof. J.L.
Vicent.

\end{document}